\documentclass[a4paper,12pt]{article}

\usepackage{pstricks}
\usepackage{amsmath}\usepackage{amssymb, graphics}
\usepackage{latexsym}\usepackage{cite}

\newcommand{\be}{\begin{equation}}
\newcommand{\ee}{\end{equation}}
\newcommand{\tr}{\,\textup{tr}}
\newcommand{\ka}{\kappa}

\def\beq{\begin{equation}}
\def\eeq{\end{equation}}

\def\al{\alpha}
\def\bt{\beta}
\def\Ga{\Gamma}

\def\de{\delta}
\def\De{\Delta}

\def\ka{\kappa}
\def\si{\sigma}

\def\te{\theta}

\def\lam{\lambda}

\def\om{\omega}
\def\ep{\epsilon}

\def\sq{\sqrt}

\def\l{\left (}
\def\r{\right )}

\def\fr{\frac}
\def\la{\label}
\def\hs{\hspace}
\def\vs{\vspace}

\def\ran{\rangle}
\def\lan{\langle}
\def\ov{\overline}
\def\tl{\tilde}
\def\tm{\times}
\def\lrar{\leftrightarrow}

\begin{document}

\begin{flushright}
OSU-HEP-08-09\\
December 23, 2008
\end{flushright}

\vs{0.5cm}

\begin{center}
{\Large\bf Common Origin for CP Violation in Cosmology \\[0.1in]
and in Neutrino Oscillations
}
\end{center}

\vspace{0.5cm}
\begin{center}
{\Large
{}~K.S. Babu,
{}~Yanzhi Meng,
{}~Zurab Tavartkiladze
}
\vspace{0.5cm}

{\em Department of Physics, Oklahoma State University, Stillwater, OK 74078, USA }

\end{center}
\vspace{0.6cm}

\begin{abstract}


We suggest predictive scenarios for neutrino masses which provide a common origin for CP violation in early universe cosmology and in neutrino oscillations. Our setup is the seesaw mechanism in the context of MSSM
with two quasi--degenerate right--handed neutrinos, with baryon asymmetry generated via resonant leptogenesis.  Three
different models are found with specific textures in the Yukawa coupling matrices, each with
a single phase which controls leptogenesis and neutrino CP violation.  One model
leads to normal hierarchy of light neutrino masses and the prediction $\tan \te_{13} \simeq \sin \te_{12} \sqrt{\frac{m_2}{m_3}}$, resulting in a value of the reactor mixing angle $\theta_{13}$ very close to
the current experimental lower limit.  The other two models predict inverted hierarchical neutrino mass spectrum
with the sum rules $\sin^2\theta_{12}\simeq \frac{1}{2}-\tan\theta_{23}\sin\theta_{13}\cos\delta $
and $\sin^2\theta_{12}\simeq \frac{1}{2}+\cot \te_{23}\sin\theta_{13}\cos\delta $ respectively.
We obtain a lower bound for the phase $|\de |$  in the normal hierarchical model, and a narrow range
for $|\de|$ for the inverted hierarchical model from cosmology. In our scenario, the mass--splitting between
the quasi--degenerate right--handed neutrinos arise via renormalization group flow, which provides a lower
limit on the MSSM parameter $\tan \bt > 12$. The right--handed neutrino masses can be as low as TeV, which
would avoid the  gravitino problem generic to supersymmetric models.

\end{abstract}

\vs{0.7cm}

\newpage

\section{ Introduction}

While the standard model (SM) of strong and electroweak interactions has been extremely successful in
confronting experimental data, it leaves several questions unanswered.  On the observational
side, it does not provide a viable dark matter candidate, nor a dynamical mechanism to explain the
observed baryon excess in the universe.  Furthermore, the model needs to be extended, albeit in a
minor way, to accommodate small neutrino masses as needed for atmospheric \cite{Fukuda:2000np} and solar neutrino oscillation data \cite{Fukuda:2001nj}. On the theoretical side, the model suffers from the quadratic divergence problem, which  destabilizes the Higgs boson mass.

An elegant synthesis of these issues is provided by low energy supersymmetry (SUSY) and the seesaw mechanism \cite{seesaw}. Low energy SUSY can cure the quadratic divergence problem for the Higgs boson mass.  In its
simplest form it also provides a natural dark matter candidate, the lightest SUSY particle (LSP).  The seesaw
mechanism assumes the existence of right--handed neutrinos (RHN) $N$ which facilitates small neutrino masses.
It also provides a dynamical mechanism for baryon asymmetry generation via the lepton number violating
decays of the $N$  \cite{Fukugita:1986hr}.  The induced lepton asymmetry is converted into
baryon asymmetry via the electroweak sphalerons \cite{Kuzmin:1985mm} (for reviews of leptogenesis see Ref.
\cite{Giudice:2003jh, Buchmuller:2004nz}).

Attractive as it is, the SUSY seesaw framework is not without its problems.  First, the generic leptogenesis
mechanism is impossible to test experimentally.  This is primarily because the dynamics occurs at a very high energy
scale, beyond reach of foreseeable experiments.  The parameters that are relevant for leptogenesis are not the
same that appear in low energy neutrino oscillation experiments.  (The number of low energy observables in
neutrino sector is nine, while leptogeneis in the general setting involves a total of eighteen parameters.)
Second, in supergravity models, successful leptogenesis is in conflict with the gravitino abundance.
This is because of the lower bound on the lightest RHN mass $M_{N_1}\stackrel{>}{_\sim}10^9$~GeV, (assuming hierarchical masses for $N$) \cite{Davidson:2002qv} which
would suggest rather high reheat temperature, of order $10^9$ GeV.  This conflicts with reheat temperature
suggested by gravitino abundance $T_{\rm reheat} < 10^8$ GeV \cite{Khlopov:1984pf,Kohri:2005wn}.

In this paper we suggest a scenario where the aforementioned problems of the SUSY seesaw framework
are alleviated.  The gravitino overproduction problem is avoided by resorting to resonant leptogenesis
scenario \cite{Flanz:1996fb, Pilaftsis:1997jf, Pilaftsis:2003gt} which assumes quasi--degenerate $N$ fields.
In this case the mass of the $N$ fields can be as low as a TeV, consistent with successful leptogenesis,
thus avoiding the gravitino problem.  We supplement the resonant leptogenesis scenario with
flavor symmetries which restrict the form of the neutrino Yukawa coupling matrices.  Such flavor symmetries
are anyway needed to guarantee the near degeneracy of the $N$ states.  We identify three possible textures
for the Dirac Yukawa couplings of the neutrinos that yield two quasi--degenerate $N$ fields and
a sum rule for the neutrino oscillation angle $\theta_{13}$.  Interestingly, in all three models, there is a
single phase that controls cosmological CP asymmetry and CP violation in neutrino oscillations.  We are
able to constrain the range of the CP violation parameter $|\delta|$ from cosmology.
Somewhat similar classification of textures has been recently pursued in Ref. \cite{Goswami:2008rt} and earlier in Ref. \cite{Frampton:2002yf}, \cite{Barbieri:1998jc}.  Our emphasis is on the connection between cosmological
CP asymmetry and CP violation in neutrino oscillations.  It turns out that, in our framework, there is a
lower limit on the SUSY parameter $\tan\beta > 12$.  This arises since the mass splitting between the
qusi--degenerate $N$ fields is generated from renormalization group flow, which depends on $\tan\beta$.

In our analysis we use the results of a global fit to the neutrino oscilation data \cite{Schwetz:2008er}:
$$
|\Delta m_{\rm atm}^2|=(2.18-2.64)\tm 10^{-3} {\rm eV}^2\,(2\si )~, \hspace{1cm}
\Delta m_{\rm sol}^2=(7.25-8.11)\tm 10^{-5} {\rm eV}^2\,(2\si )~,
$$
\beq
\sin^2\theta_{23}=0.39-0.63\, (2\sigma)~, \hspace{2.5cm}
\sin^2\theta_{12}=0.27-0.35 \,(2\sigma)~.
\la{data1}
\eeq
Currently we do not know the sign of $\Delta m_{\rm atm}^2$, i.e. whether neutrinos have normal mass hierarchy
or inverted mass hierarchy. Also, the value of the third mixing angle $\te_{13}$ is unknown. Only an upper bound \cite{Schwetz:2008er}
\begin{equation}
\sin^2\theta_{13}\leq 0.04 \,(2\sigma)
\la{data2}
\eeq
\vspace*{-0.1in}

\noindent
is available currently. Nothing is known about the CP violating phase $\de $ (and also about two `Majorana' phases) of the leptonic mixing matrix.

We will identify explicit models wherein these unknown mixing
parameters are significantly constrained.  It will be highly desirable to relate the
CP violation parameters in the leptonic mixing matrix with the cosmological CP asymmetry.
Such a strategy was pursued successfully in  Ref. \cite{Frampton:2002qc}. While in Ref. \cite{Frampton:2002qc} a close connection between cosmological CP violation and neutrino CP violation was realized, since the setup used
hierarchical RHN masses, straightforward SUSY extension of that
scenario would lead to gravitino overproduction. Our texture models are tailor--made for resonant leptogenesis,
which would avoid this problem.\footnote{For a concrete demonstration  within predictive model see \cite{Babu:2007zm}.}

\section{Texture Zeros for Predictive Models}

Let us consider the lepton sector of MSSM augmented with two right--handed neutrinos (RHN) $N_1$ and $N_2$.
The relevant Yukawa superpotential couplings are given by
$$
W_{\rm lept}=W_e+W_{\nu }~,
$$
\beq
{\rm with}~~~W_e=l^TY_ee^ch_d~,~~~~W_{\nu }=l^TY_{\nu }Nh_u-\fr{1}{2}N^TM_NN~,
\la{Wlept}
\eeq
where $h_d$ and $h_u$ are up and down type  MSSM Higgs doublet superfields respectively. We will work in a basis in which the charged lepton Yukawa matrix is diagonal:
\beq
Y_e={\rm Diag}\l \lam_e, ~\lam_{\mu }, ~\lam_{\tau }\r ~.
\la{Ye}
\eeq
As far as the RHN mass matrix $M_N$ is concerned, we will assume that at high scale (identified with the GUT scale later on)
it has the form
\beq
M_N=\left( \begin{array}{cc} 0&1\\1&0 \end{array}\right)M~.
\la{MN}
\eeq
This form of $M_N$ is crucial for our studies. It has interesting implications for resonant leptogenesis and also, as we will see below,
for building predictive neutrino scenarios. Specific neutrino models consistent with resonant leptogenesis with a texture similar to (\ref{MN})
was investigated in \cite{Babu:2007zm}. Here we attempt to classify  all possible scenarios with degenerate RHNs which lead to predictions
consistent with experiments. Thus, with a basis (\ref{Ye}) and the texture (\ref{MN}) we can discuss possible texture zeros in the
matrix $Y_{\nu }$, which is of dimension $3\tm 2$. One can easily verify that two (and more) texture zeros in $Y_{\nu }$ do not lead to results
compatible with the neutrino data. However, with only one texture zero, there are  scenarios compatible with experiments and leading to interesting predictions.

The matrix $Y_{\nu }$ contains two columns. Since due to the form of $M_N$ (\ref{MN}) there is exchange invariance $N_1\to N_2$, $N_2\to N_1$, it does not
matter in which column of $Y_{\nu }$ we set one element to zero. We choose here the second column of $Y_{\nu }$ having one texture zero.
This leads to the three following possible forms for $Y_{\nu }$:
\beq
{\rm Texture~ A:}~~~~~~~~~Y_{\nu}=\l \!\begin{array}{cc}
a_1&0\\
a_2&b_2\\
a_3&b_3
  \end{array}\!\r  ~,
\la{textureA}
\eeq
\beq
{\rm Texture~ B_1:}~~~~Y_{\nu}=\l \!\begin{array}{cc}
a_1&b_1\\
a_2&0\\
 a_3&b_3
  \end{array}\!\r  ~,~~~~~~~
{\rm Texture~ B_2:}~~~~Y_{\nu}=\l \!\begin{array}{cc}
a_1&b_1\\
a_2&b_2\\
a_3&0
  \end{array}\!\r ~.
\la{textureB12}
\eeq
A few words about the parametrization, used in (\ref{textureA}) and (\ref{textureB12}), are in order. With the basis (\ref{Ye}) and the form of $M_N$
given in (\ref{MN}), the one texture zero $3\tm 2$ matrix $Y_{\nu }$ has only one physical phase. Other phases can be rotated away by proper
phase redefinitions of the fields. Moreover, in $Y_{\nu }$ there are five real parameters $|a_{1,2,3}|$ and two
absolute values of the $b$-entries. The mass parameter $M$ in (\ref{MN}) is in general complex,
but its phase is not relevant for the physics of  neutrino oscillations.
These systems lead to predictive scenarios with texture $A$ corresponding to normal mass hierarchy and
textures $B_1$ and $B_2$ corresponding to inverted mass hierarchy. We will study these cases in turn.

\subsection{Texture A:~Normal Hierarchical Case}

We will discuss this case in details. With (\ref{MN}), (\ref{textureA}) and using the seesaw formula for the light neutrino mass
matrix $M_{\nu }=\lan h_u^0\ran^2Y_{\nu }M_N^{-1}Y_{\nu }^T$, we arrive at
\beq
M_{\nu}=\l \!\begin{array}{ccc}
0&a_1b_2&a_1b_3\\
a_1b_2&2a_2b_2&a_2b_3+a_3b_2\\
a_1b_3& a_2b_3+a_3b_2 &2a_3b_3\end{array}\!\r \!\fr{(v\sin \bt )^2}{M}~,
\la{Mnu-A}
\eeq
where $v\simeq 174$~GeV. The matrix in (\ref{Mnu-A}) is rank two and leads to the one massless neutrino and two massive
neutrinos labeled  $m_2$ and $m_3$. This structure corresponds to the normal hierarchical case, i.e.
\beq
M_{\nu}^{\rm diag}={\rm Diag}\l 0 ,~ m_2 ,~ m_3\r~,
\la{Mnu-Diag-A}
\eeq
with $m_3\gg m_2$. From (\ref{Mnu-A}) we can see that the mixings $\te_{12}$ and $\te_{23}$ are generated.
The absolute value of the overall factor  $a_3b_3(v\sin \bt )^2/M$ determines one mass scale, say the value of $m_3$.
Besides this overall factor the matrix has four  parameters: one phase and three real parameters.
Three of these parameters can be fixed from  three observables
$\te_{12}$, $\te_{23}$ and $\fr{\De m_{\rm sol}^2}{\De m_{\rm atm}^2}$ (where $\De m_{\rm sol}^2=m_2^2$ and $\De m_{\rm atm}^2=m_3^2-m_2^2$).
Due to the condition $m_1=0$ we will still have one prediction (independent from the value of the phase),
which determines the angle $\te_{13}$.

One physical phase  remains undetermined. Indeed this {\it single} phase will be directly related to the CP violation
in neutrino oscillations and in leptogenesis. We will discuss this connection in more details in Sect. \ref{s:res}.

Now, let us derive the prediction of this model. To achieve this and also get other useful relations we will use the equality
\begin{equation}
M_{\nu}=PU^{*}P'M_{\nu}^{\rm diag}U^{\dagger}P,
\la{relM-Md}
\end{equation}
where $U$ is the lepton mixing matrix, given in a standard parameterization by:
\begin{equation}
U=
\left(\begin{array}{ccc}
c_{13}c_{12}&c_{13}s_{12}&s_{13}e^{-i\delta}\\
-c_{23}s_{12}-s_{23}s_{13}c_{12}e^{i\delta}&c_{23}c_{12}-s_{23}s_{13}s_{12}e^{i\delta}&s_{23}c_{13}\\
s_{23}s_{12}-c_{23}s_{13}c_{12}e^{i\delta}&-s_{23}c_{12}-c_{23}s_{13}s_{12}e^{i\delta}&c_{23}c_{13}
\end{array}\right)~,
\la{Ustan}
\end{equation}
with $s_{ij}=\sin\theta_{ij}$ and $c_{ij}=\cos\theta_{ij}$.
The $P$ and $P'$ are diagonal phase matrices $P={\rm Diag}\l e^{i\om_1} ,e^{i\om_2} , e^{i\om_3} \r$, $P'={\rm Diag}\l 1 ,e^{i\rho_1} , e^{i\rho_2} \r$. Phases in $P$ can be removed by field redefinition, while $P'$ is physical, and
contains the two Majorana phases.  The matrix equation (\ref{relM-Md}) gives six relations. One
of them, namely the relation for the $(1,1)$ elements of $M_{\nu}$ and the right hand side of (\ref{relM-Md}) with the form of $U$ given in Eq. (\ref{Ustan}),
gives
\begin{equation}
\tan\theta_{13}\simeq \sin\theta_{12}\sqrt{\frac{m_2}{m_3}}~.
\la{predA}
\end{equation}
Since this case corresponds to the normal hierarchical neutrino mass spectrum (with $m_1=0$), with the help of (\ref{data1}) we have at $2\si $
level $m_2=\sq{\De m_{\rm sol}^2}\simeq (8.51-9.01)\cdot 10^{-3}$~eV and $m_3=\sq{|\De m_{\rm atm}|^2+\De m_{\rm sol}^2}\simeq (4.7-5.2)\cdot 10^{-2}$~eV.
Using these values in (\ref{predA}), together with $2\si $ accuracy value of
$\te_{12}$, we obtain range $\sin^2\te_{13}\simeq 0.042-0.062$. This fits well with an upper bound, within $3\si $, given in
Ref. \cite{Schwetz:2008er}, while the low limit $(0.042)$ is pretty close to the $2\si $ upper bound of $\te_{13}$.
Future measurements of the $\te_{13}$ will test the validity of this scenario.
One more word about the neutrino sector: since the $(1,1)$ element of the light neutrino mass matrix vanishes,
the neutrino--less double $\beta $-decay ($0\nu 2\bt $) does not take place in this scenario. That is,
$m_{\bt \bt }=|U_{e2}^2m_2e^{i\bar{\rho }}+U_{e3}^2m_3|=0$. There is only one Majorana phase, since $m_1 = 0$, which
is  $\bar{\rho }=\rho_2-\rho_1$.  This is determined from the phase $\de $ as follows
\beq
\bar{\rho }=\pi-2\de ~.
\la{A-maj-phase}
\eeq

\subsection{Textures $B_1$ and $B_2$~:~ Inverted Hierarchical Cases}

The textures $B_1$ and $B_2$ both lead to the inverted hierarchical neutrino mass pattern.
Using these textures (\ref{textureB12}), the form of $M_N$ given in (\ref{MN}) and the seesaw formula for the light neutrino mass matrices we obtain:
\beq
{\rm For~ Texture~B_1} :~~~~M_{\nu}=\l \!\begin{array}{ccc}
2a_1b_1& a_2b_1& a_1b_3+a_3b_1\\
a_2b_1&0&a_2b_3 \\
a_1b_3+a_3b_1 & a_2b_3 &2a_3b_3\end{array}\!\r \!\fr{(v\sin \bt )^2}{M}~,
\la{Mnu-B1}
\eeq
\beq
{\rm For~ Texture~B_2} :~~~~M_{\nu}=\l \!\begin{array}{ccc}
2a_1b_1 &a_1b_2+a_2b_1&a_3b_1\\
a_1b_2+a_2b_1 &2a_2b_2& a_3b_2\\
a_3b_1& a_3b_2 &0\end{array}\!\r \!\fr{(v\sin \bt )^2}{M}~.
\la{Mnu-B2}
\eeq
In order to derive predictions for both cases, we can still use the relation (\ref{relM-Md}), which is general, but for $M_{\nu }$
use the forms corresponding to the cases $B_{1,2}$, and for $M_{\nu}^{\rm diag}$ an inverted hierarchical form:
\beq
M_{\nu}^{\rm diag}={\rm Diag}\l m_1 ,~ m_2 ,~ 0\r~.
\la{Mnu-Diag-B}
\eeq
We use the same form as before for the phase matrix $P$, while for the  $P'$ we use $P'={\rm Diag}\l e^{i\rho_1}, e^{i\rho_2},1\r$.
For cases $B_1$ and $B_2$ the predictive relations emerge by equating the $(2,2)$ and $(3,3)$ elements respectively (which are zero)
of the expressions at the both sides of Eq. (\ref{relM-Md}). Doing so we arrive at:
\beq
{\rm For~ texture~ B_1} :~~~~
\sin^2\theta_{12}\simeq\frac{1}{2}-\frac{\sin\theta_{13} \tan \theta_{23}\cos\delta}{|\tan^2 \theta_{23}\sin^2 \theta_{13}+e^{2i\delta}|}+
\fr{1}{8}\fr{\De m_{\rm sol}^2}{|\De m_{\rm atm}^2|}~,
\la{predB1}
\eeq
\beq
{\rm For~ texture~ B_2} :~~~~
\sin^2\theta_{12}\simeq\frac{1}{2}+\frac{\sin\theta_{13} \tan \theta_{23}\cos\delta}{|\tan^2 \theta_{23}+\sin^2 \theta_{13}e^{2i\delta}|}+
\fr{1}{8}\fr{\De m_{\rm sol}^2}{|\De m_{\rm atm}^2|}~.
\la{predB2}
\eeq
As we see, for both cases, the deviation of $\sin^2\theta_{12}$ from $1/2$ (i.e. deviation of $\te_{12}$ from $\pi/4$) is due to
the non--zero value of $\te_{13}$ and it also depends on $\cos \de $\footnote{Similar relation has been obtained in Ref. \cite{Babu:2007zm} within a
specific model with $\te_{23}\simeq \pi/4$. Here, since $\te_{23}$ is not fixed from the model, we will have somewhat wider allowed ranges for
$\te_{13}$ and especially for $\de $. Cases of texture zeros giving these relations have been identified recently in Ref. \cite{Goswami:2008rt}.
Correlation similar to Eqs. (\ref{predB1}) and (\ref{predB2}) have been obtained within
scenarios with `quark-lepton complementarity' \cite{Giunti:2002ye}.}.
In fact, the product $\sin \te_{13}\cos \de $ should not be too small, otherwise the angle $\te_{12}$ will be close to $\pi/4$ which is excluded.
Using the current experimental data (\ref{data1}) (within $2\si$-deviations) we obtain the following constraints for $\te_{13}$ and $\cos \de $:
$$
{\rm For~ texture~ B_1} :~~~~\te_{13}\stackrel{>}{_\sim}0.12~,~~~~~\cos \de \stackrel{>}{_\sim}0.573~~~(|\de |\stackrel{<}{_\sim}0.96)~,
$$
\beq
{\rm For~ texture~ B_2} :~~~~\te_{13}\stackrel{>}{_\sim}0.129~,~~~~~\cos \de \stackrel{<}{_\sim}-0.614~~~(|\pi -\de |\stackrel{<}{_\sim}0.91)~.
\la{constr-13-del-B12}
\eeq
The last terms in Eqs. (\ref{predB1}), (\ref{predB2}) are practically unimportant for the neutrino sector, but as we will
see in section \ref{lep-gen-fermion} they become crucial for the leptogenesis CP violation.
The leptonic asymmetry will be determined by a phase $\propto \fr{\De m_{\rm sol}^2}{|\De m_{\rm atm}^2|}\sin \de $
which would vanish in the limit $\fr{\De m_{\rm sol}^2}{|\De m_{\rm atm}^2|}\to 0$.

By the fixed  model parameters (see sect. \ref{lep-gen-fermion} for relation between Yukawa couplings and the angles $\te_{ij}, \de $) we can compute one more observable. In contrast to the normal hierarchical
neutrinos (corresponding to the texture A), cases $B_1$ and $B_2$ have non--zero
$\beta\beta_{0\nu}$ amplitudes, for both cases given by
\beq
m_{\bt \bt}=|U_{e1}^2m_1e^{i\bar{\rho}}+U_{e2}^2m_2|~,~~~~{\rm with}~~~~\bar{\rho}=\rho_2-\rho_1~.
\la{B12-gen-2bt}
\eeq
For $m_{\bt \bt}$ and $\bar{\rho}$ for scenarios $B_1$ and $B_2$  respectively we derive:
$$
{\rm For~ texture~ B_1} :~~~~~~~~m_{\bt \bt}=\sq{|\De m_{\rm atm}^2|} c_{13}^2
\fr{|{\rm tg_{12}^2-1+2tg_{12}tg_{23}}s_{13}e^{i\de }|}
{|{\rm tg_{12}+tg_{23}}s_{13}e^{i\de }|^2}~,
$$
\beq
\cot \fr{\bar{\rho}}{2}=-
\fr{{\rm tg_{23}(1+tg_{12}^2)}s_{13}\sin \de }{{\rm tg_{12}(1-tg_{23}^2}s_{13}^2)+{\rm tg_{23}(1-tg_{12}^2)}s_{13}\cos \de }~,
\la{B1-anal-mbb-rho}
\eeq

$$
{\rm For~ texture~ B_2} :~~~~~~~~m_{\bt \bt}=\sq{|\De m_{\rm atm}^2|} c_{13}^2
\fr{|{\rm tg_{12}^2-1-2tg_{12}ctg_{23}}s_{13}e^{i\de }|}
{|{\rm tg_{12}-ctg_{23}}s_{13}e^{i\de }|^2}~,
$$
\beq
\cot \fr{\bar{\rho}}{2}=
\fr{{\rm tg_{23}(1+tg_{12}^2)}s_{13}\sin \de }{{\rm tg_{12}(tg_{23}^2}-s_{13}^2)-{\rm tg_{23}(1-tg_{12}^2)}s_{13}\cos \de }~,
\la{B2-anal-mbb-rho}
\eeq
where ${\rm tg_{ij}}\equiv \tan \te_{ij}$ and ${\rm ctg_{ij}}\equiv \cot \te_{ij}$. Applying allowed ranges for $\de $ and $\te_{13}$
given in Eq. (\ref{constr-13-del-B12}) and the measured neutrino oscillation parameters (\ref{data1}) (within $2\si $) for $m_{\bt \bt }$ we obtain:
\beq
{\rm For~ textures~ B_1~\& ~B_2} :~~~~ 0.013~{\rm eV} \stackrel{<}{_\sim }m_{\bt \bt} \stackrel{<}{_\sim }0.023~{\rm eV}~.
\la{bt-decay-B12}
\eeq
Upper bounds for $m_{\bt \bt}$ are obtained for $|\de |=0.96$ and $|\pi -\de |=0.91$ for cases $B_1$ and $B_2$ respectively,
while lower limits correspond to $\de =0$ and $\de =\pm \pi$.
Planned experiments will certainly be able to test viability of these predictions.
Note that the textures $B_1$ and $B_2$ in the  neutrino sector give results which are practically indistinguishable (besides the allowed ranges
for $\de $).
However, as we will see in the next section the scenario $B_2$ fails to generate sufficient leptogenesis, while the
texture $B_1$ (and also the texture A) will work very well for this purpose.

\section{Resonant Leptogenesis}
\la{s:res}

Within the scenarios considered in the previous section, we have assumed an off--diagonal form for the RHN mass
matrix $M_N$. This gives the desired degeneracy between the two RHN states. The degeneracy will be lifted with
small corrections to the $(1,1)$ and/or $(2,2)$ elements of $M_N$. Even in the unbroken SUSY limit, 1-loop corrections (corresponding
to the wave function renormalization) will split the  degeneracy. The SUSY breaking effects has dramatic impact on the degeneracy of the scalar components of
$N_{1,2}$ superfields. This is discussed separately in the Appendix. As far as the fermionic RHN sector is concerned, the degeneracy there holds with pretty high accuracy. Therefore, this is an appealing framework for
resonant leptogenesis, in which enhancement of the CP asymmetry happens because of
quasi-degenerate RHN neutrinos \cite{Flanz:1996fb, Pilaftsis:1997jf, Pilaftsis:2003gt}. One nice property of the resonant leptogenesis is that, it avoids the lower bound
($M_{N_1}\stackrel{>}{_\sim }10^9$~GeV) for the lightest RHN mass. This bound, called as Davidson-Ibarra bound,
emerges within most of the scenarios with hierarchical right--handed neutrinos \cite{Davidson:2002qv}.
Once this bound is avoided, the reheat temperature can be sufficiently low to avoid the gravitino problem, which
is common for low scale SUSY models  \cite{Khlopov:1984pf, Kohri:2005wn}  with the gravity mediated SUSY breaking.

Since our models of neutrino masses and mixings are predictive and involve very limited number of parameters,
we expect that we will not have much freedom in the calculation of leptogenesis.
As we have already mentioned, an important ingredient for the resonant leptogenesis is the form of $M_N$ given in Eq. (\ref{MN}).
Note that the mass matrix of the fermionic RHNs coincides with $M_N$ of the superpotential mass term. First we will discuss
radiative corrections to the superpotential mass matrix $M_N$, which directly can be applied to the fermionic RHNs.
This structure can be justified by some symmetry at high scale. However, at low energies, due to the radiative corrections the $(1,1)$
and $(2,2)$ entries in $M_N$ will receive non-zero corrections. These corrections  are calculable thanks
to the well defined neutrino models we have presented above. To be brief, eventually two RHNs are become quasi-degenerate, and the CP asymmetries $\ep_1$ and $\ep_2$
generated by out-of-equilibrium decays of the fermionic components of $N_1$ and $N_2$ states respectively are given by
\cite{Pilaftsis:1997jf, Pilaftsis:2003gt}
\beq
\ep_1=\fr{{\rm Im}(\hat{Y}_{\nu }^{\dagger}\hat{Y}_{\nu })_{21}^2}
{(\hat{Y}_{\nu }^{\dagger}\hat{Y}_{\nu })_{11}(\hat{Y}_{\nu }^{\dagger}\hat{Y}_{\nu })_{22}}
\fr{\l M_2^2-M_1^2\r M_1\Ga_2}{\l M_2^2-M_1^2\r^2+M_1^2\Ga_2^2}~,
~~~~~~~~~~~~\ep_2=\ep_1(1\lrar 2)~.
\la{res-lept-asym}
\eeq
Here $M_1$ and $M_2$ (we will use the convention $M_2>M_1$) are the mass eigenvalues of the RH neutrino mass eigenstates.
$\hat{Y}_{\nu }=Y_{\nu}U_N$ is the Dirac Yukawa coupling matrix in the basis where RH neutrino mass matrix is diagonal and
real: $U_N^TM_NU_N={\rm Diag}\l M_1 ,~ M_2\r$. $\Ga_i$ is the tree level decay width of $\bar N_i$ (mass eigenstates of RHN)
and is given by $\Ga_i=\fr{M_i}{4\pi}(Y_{\nu}^{\dag}Y_{\nu})_{ii}$. From (\ref{res-lept-asym}) we see that in order to have non--zero CP asymmetry
two conditions need to be satisfied. First, the RHN masses should be split, and secondly the element $(\hat{Y}_{\nu}^{\dag}\hat{Y}_{\nu})_{12}$
must be complex. To realize both of these conditions, we need to include radiative corrections into our study.  As we will see shortly, the desired
result can be obtained only at two-loop level. In our treatment we assume that the textures we have considered
are realized at the GUT scale $M_G\simeq 2\cdot 10^{16}$~GeV. At low scales, due to the renormalization group
effects the zero entries in the
flavor matrices will receive some corrections. To compute these corrections we set up the RG equation for the matrix $M_N$
(only its renormalization has relevance for us), which at two--loop order is given by \cite{Antusch:2002ek}:
$$
16\pi^2\fr{d}{dt}M_N=2M_NY_{\nu}^{\dag}Y_{\nu}+2Y_{\nu }^TY_{\nu }^*M_N
$$
$$
-\fr{1}{8\pi^2}M_N\l Y_{\nu}^{\dag}Y_eY_e^{\dag}Y_{\nu}+Y_{\nu}^{\dag}Y_{\nu}Y_{\nu}^{\dag}Y_{\nu}+Y_{\nu}^{\dag}Y_{\nu}(3\lam_t^2+{\tr}(Y_{\nu}^{\dag}Y_{\nu})\r
$$
\beq
-\fr{1}{8\pi^2}\l Y_{\nu}^TY_e^*Y_e^TY_{\nu}^* +Y_{\nu}^TY_{\nu}^*Y_{\nu}^TY_{\nu}^*+Y_{\nu}^TY_{\nu}^*(3\lam_t^2+{\tr}(Y_{\nu}^{\dag}Y_{\nu})\r M_N+\fr{3}{20\pi^2}M_N\l g_1^2+5g_2^2\r~,
\la{MN-2loop-RG}
\eeq
where $t=\ln \mu $.
The first line in (\ref{MN-2loop-RG}) corresponds to the 1-loop correction and will be responsible for the mass splitting between RHNs. However,
the two-loop correction, presented in  a second line of Eq. (\ref{MN-2loop-RG}), will be crucial for the CP phase of $(\hat{Y}_{\nu}^{\dag}\hat{Y}_{\nu})_{12}$.
Since we intend to have $M_{1,2}\stackrel{<}{_\sim}10^7$~GeV, in order to get reasonable scale for the light neutrino mass, the matrix elements
of $Y_{\nu}$ should be much less than unity. Thus, we can solve the RG equation analytically to a good approximation.
One--loop correction to the $M_N$ can be found from (\ref{MN-2loop-RG}) to be
\beq
\de M_N^{\rm 1-loop}\simeq -\fr{1}{8\pi^2}\l M_NY_{\nu}^{\dag}Y_{\nu}+Y_{\nu}^TY_{\nu}^*M_N\r_{\mu =M_G}\ln \fr{M_G}{M}~.
\la{sol-1loop}
\eeq
From this we see that at scale $\mu =M$ the form of $M_N$ will become
\beq
 M_N =M \l \begin{array}{cc}-\delta_{N}&1\\1&-\delta^*_N\end{array}\r~.
\la{effectiveMN}
\eeq
Interestingly enough, this structure, of correlated phases of $(1,1)$ and (2,2) entries of $M_N$, persists also at two--loop order. What is
more important, one can see that at the one--loop level the phase of $\de_{N}$ is determined by the phase of $(Y_{\nu}^{\dag}Y_{\nu})_{12}$
and therefore $(\hat{Y}_{\nu}^{\dag}\hat{Y}_{\nu})_{12}$ will be real at this level. This property can be easily seen also from different angle. Regardless of the form of $Y_{\nu}$ (including all possible radiative corrections to it), it can be written in the form
\beq
Y_{\nu}={\cal U}\l \!\begin{array}{cc}
0&0\\
\hat{a}_2&0\\
 \hat{a}_3&\hat{b}_3
  \end{array}\!\r \!\tl{P} ~,
\la{realBasis}
\eeq
with ${\cal U}$ some unitary matrix, $\hat{a}_{1,2}, \hat{b}_3$ all real parameters and $\tl{P}={\rm Diag}\l 1, e^{i\xi }\r$. Using now the form (\ref{realBasis})
in the first line of (\ref{MN-2loop-RG}), one can show that ${\cal U}$ drops out and we remain with the non physical phase $\xi $ which can be absorbed in
$N_2$. With this, any complexity in $\de M_N^{\rm 1-loop}$ and  $(Y_{\nu}^{\dag}Y_{\nu})_{12}$ disappears and we have no CP violation at the one--loop
level. That is why it is important to include two--loop radiative corrections for the renormalization of $M_N$. Indeed, the term
$M_NY_{\nu}^{\dag}Y_eY_e^{\dag}Y_{\nu}$
in the second
line of Eq. (\ref{MN-2loop-RG}) is important. The appearance of the combination $Y_eY_e^{\dag}$ plays an important role. With the basis (\ref{realBasis})
we see that the matrix ${\cal U}$ does not disappear, and thus we expect to have CP violation (induced through two--loop correction). From (\ref{MN-2loop-RG}), this correction can be approximated as follows:
\beq
\de M_N^{\rm 2-loop}\simeq \fr{2}{(16\pi^2)^2}\l M_NY_{\nu}^{\dag}Y_eY_e^{\dag}Y_{\nu}+Y_{\nu}^TY_e^*Y_e^TY_{\nu}^*M_N\r_{\mu =M_G}R_{\ell }
\ln \fr{M_G}{M}~.
\la{sol-2loop}
\eeq
where we have suitably absorbed CP conserving and flavor universal corrections (coming from the entries ${\rm Tr}(Y_{\nu}^{\dag}Y_{\nu})$, $g_i^2$, $\lam_t^2$ etc.)
in the overall scale $M$. The RG factor $R_{\ell }$ ($\ell_i=(e,\mu ,\tau )$) is for the running of $Y_e$ Yukawa couplings and can strongly deviate from one.
It is defined as:
\beq
R_{e, \mu ,\tau}=\fr{\int_{M}^{M_G}\lam^2_{e,\mu ,\tau }(t)dt}{\lam^2_{e, \mu ,\tau}(M_Z)\ln \fr{M_G}{M}}~.
\la{RtauDef}
\eeq
In the approximation (\ref{sol-2loop}), the fourth powers of $Y_{\nu}$ have been neglected.
Actually, for calculating the mass splitting $M_2^2-M_1^2$ - the combination appearing in (\ref{res-lept-asym}) - it is enough to keep only correction
$\de M_N^{\rm 1-loop}$ of (\ref{sol-1loop}). However, to deal with the CP violating effect, we need to include also two--loop effects. Thus, at the scale
$\mu =M$ for $M_N$ we use
\beq
M_N(M)=M_N(M_G)+\de M_N^{\rm 1-loop}+\de M_N^{\rm 2-loop}~,
\la{renormMN}
\eeq
with $M_N(M_G)$, $\de M_N^{\rm 1-loop}$ and $\de M_N^{\rm 1-loop}$ given by Eqs. (\ref{MN}), (\ref{sol-1loop}) and (\ref{sol-2loop}) respectively.
This completes the calculation of supersymmetric part, which will be useful for calculation of leptogenesis via fermionic RHN decays.
However, inclusion of soft SUSY breaking terms, in general, may affect the leptogenesis induced through the
right--handed sneutrino decays.
In an Appendix we studied this case separately and shown that under plausible assumptions the right--handed sneutrino decays practically do not contribute to the net baryon asymmetry.
Thus, we should relay on  the fermionic RHN decays which, as we show below, generate sufficient baryon asymmetry.

\subsection{Asymmetry Via Fermionic RHN Decays}
\la{lep-gen-fermion}

\begin{center}
{\bf Leptogenesis for Normal Hierarchical Case}
\end{center}

For this case we will take the form of $Y_{\nu}$ given by Eq. (\ref{textureA}). For leptogenesis study, it is convenient to parameterize this Yukawa matrix as follows:
\beq
{\rm Texture~ A:}~~~~~~~~~Y_{\nu}=\l \!\begin{array}{cc}
x\al_1&0\\
x\al_2&b\\
 xe^{i\phi}&1
  \end{array}\!\r \!\cdot \!\bar{\bt } ~,
\la{1textureA}
\eeq
where the couplings $\al_{1,2}, b, \ov{\bt }$ and $x$ are real parameters. Only single phase $\phi $ appears. This has been achieved
by suitable redefinition of phases of $l_{1,2,3}$ and $N_{1,2}$ superfields.
First we will relate the parameters appearing
in $Y_{\nu}$ to some observables.
The relation (\ref{relM-Md}) enables us to express
$\al_1, \al_2, b$ and $\bar{\bt}$ in terms of $x$, neutrino mass, the scale $|M|$ and lepton mixing matrix elements.
Also $\phi $ can be determined by the phase $\de $ and the leptonic mixing angles.
Doing so, we arrive to the following relations
\beq
\al_1=2\left | \fr{{\cal A}_2}{{\cal A}_1}\right |~,~~~\al_2=\left | \fr{{\cal A}_2{\cal A}_4}{{\cal A}_1{\cal A}_3}\right |~,~~~
b=\left | \fr{{\cal A}_3}{{\cal A}_2}\right |~,~~~
\bar{\bt }=\fr{1}{v\sin \bt}\l \fr{m_3}{2}\left | \fr{{\cal A}_1M}{x}\right |\r^{1/2}~,
\la{rel1-A}
\eeq
\beq
\phi ={\rm Arg}\l \fr{{\cal A}_2^2{\cal A}_4}{{\cal A}_1{\cal A}_3^2}\r ~,
\la{rel2-A}
\eeq
\begin{figure}[!ht]
  \begin{flushleft}
 \vs{0cm}
    \begin{tabular}{cc}\hs{0.15cm}
            \resizebox{0.46\textwidth}{!}{\includegraphics{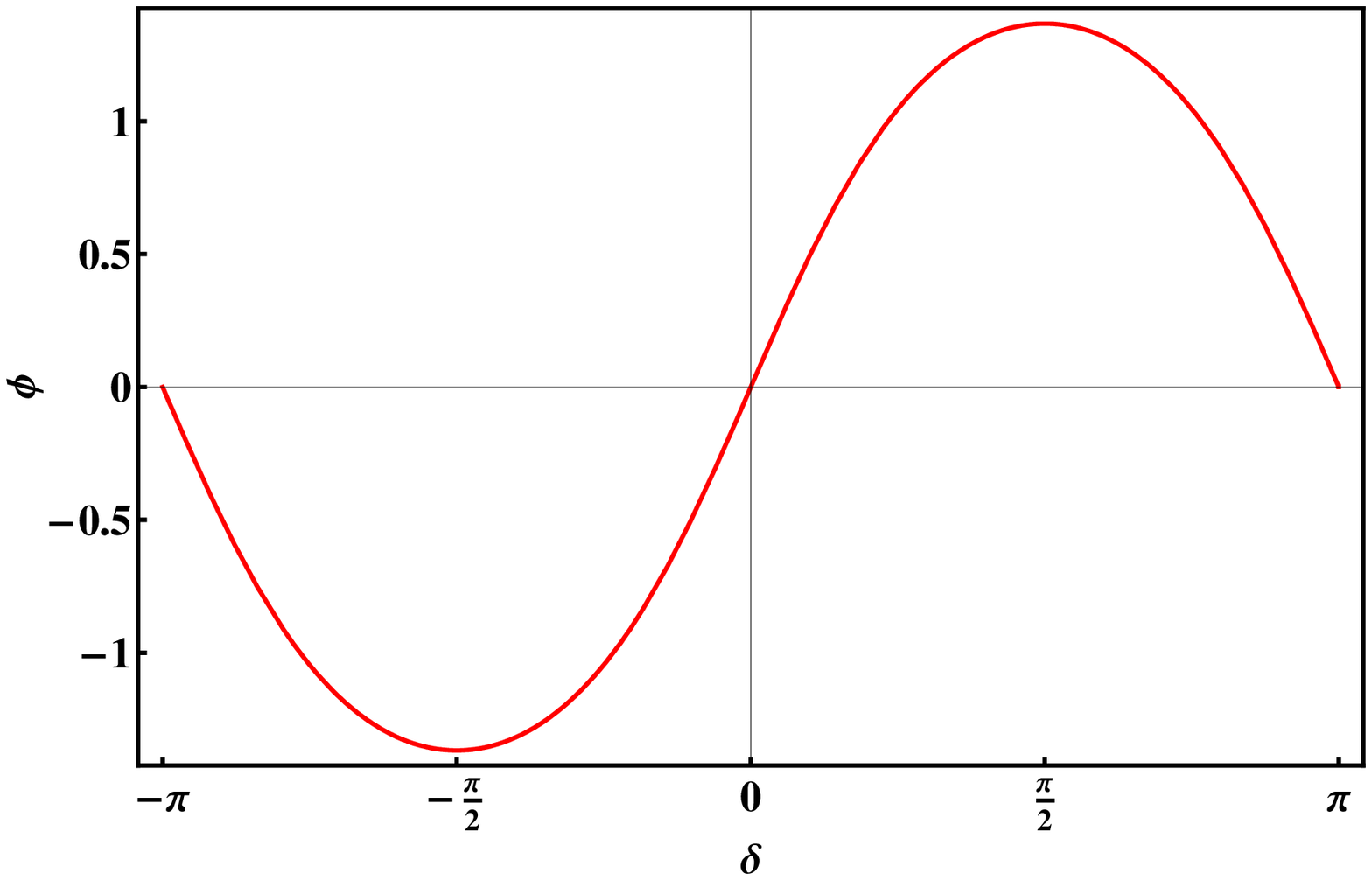}}
            &\hs{0.1cm}
      \resizebox{0.46\textwidth}{!}{\includegraphics{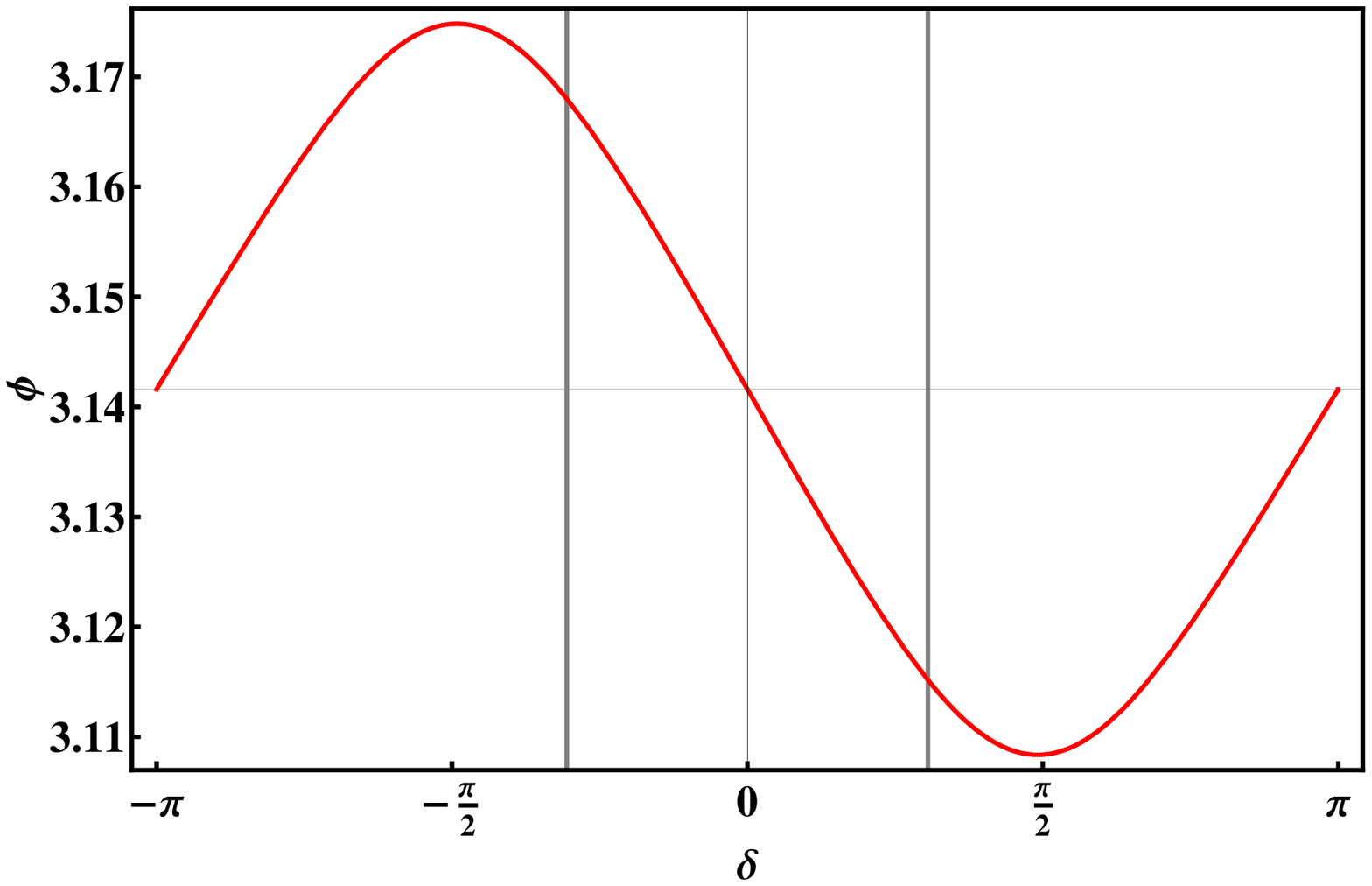}}
       \\
    \end{tabular}
    \caption{Correlation between $\phi $ and $\de $. Left side: normal hierarchical case (texture A). Right side: inverted hierarchical case
    (texture $B_1$). The vertical lines, for right panel, correspond to the maximal allowed value of $|\de |=0.96$.
    }
    \label{f:phi-del}
  \end{flushleft}
\end{figure}
where
\beq
{\cal A}_1=U_{\tau 3}^2-U_{\tau 2}^2\fr{U_{e3}^2}{U_{e2}^2}~,~~~
{\cal A}_2=U_{e3}U_{\tau 3}-U_{\tau 2}\fr{U_{e3}^2}{U_{e2}}~,~~~
{\cal A}_3=U_{e3}U_{\mu 3}-U_{\mu 2}\fr{U_{e3}^2}{U_{e2}}~,~~~
{\cal A}_4=U_{\mu 3}^2-U_{\mu 2}^2\fr{U_{e3}^2}{U_{e2}^2}~.
\la{defA1234}
\eeq
These will be useful upon studying the leptogenesis. As we have already mentioned, remarkable thing is the fact that there is a single CP violating
phase $\phi$ which is related to the phase $\de $ controlling the CP violation in the neutrino oscillations. The same phase will
appear in the CP asymmetry of the resonant leptogenesis. In Fig. \ref{f:phi-del} we show correlation between $\phi $ and $\de $.

Furthermore, applying expressions (\ref{sol-1loop}), (\ref{sol-2loop}),
 for the splitting parameter $\de_N$ of (\ref{effectiveMN}) we obtain
\beq
\de_N\simeq \l b\al_2+e^{i\phi}-\fr{\lam_{\tau}^2R_{\tau }}{16\pi^2}e^{i\phi }\r \fr{x\bar{\bt}^2}{4\pi^2}\ln \fr{M_G}{M}~,
\la{deltaN}
\eeq
where we have ignored the couplings $\lam_e$ and $\lam_{\mu }$ because the main effect is obtained by the tau Yukawa coupling.
In Eq. (\ref{deltaN}), the coupling $\lam_{\tau }$ is defined at $M_Z$ scale, and therefore the quantity $R_{\tau }$
accounts for the renormalization effects mostly due to $\lam_{\tau }$ running, and is given in Eq. (\ref{RtauDef}).
Now we can give the unitary matrix $U_N$ diagonalizing $M_N$ (by the transformation $U_N^TM_NU_N=M_N^{\rm Diag}$):
\beq
U_N \simeq
\frac{1}{\sqrt{2}}\left(\begin{array}{cc}e^{-i\eta/2}&-ie^{-i\eta/2}\\
e^{i\eta/2}&ie^{i\eta/2}\end{array}\right)~,
\la{UNform}
\eeq
where
\beq
\eta ={\rm Arg}\l b\al_2+e^{i\phi}-\fr{\lam_{\tau}^2R_{\tau }}{16\pi^2}e^{i\phi }\r~.
\la{eta}
\eeq
At the same time we have
\beq
(Y_{\nu}^{\dag}Y_{\nu})_{21}=\bar{\bt}^2x\left |b\al_2+e^{i\phi}\right |e^{i\eta'}~,~~~~~{\rm with}~~~~~
\eta'={\rm Arg}\l b\al_2+e^{i\phi}\r ~.
\la{Y2-21el}
\eeq
Therefore, the complex phase appearing in $(\hat{Y}_{\nu}^{\dag}\hat{Y}_{\nu})_{21}$ will be proportional to the mismatch $\eta -\eta'$,
which using (\ref{eta}) and (\ref{Y2-21el}) takes the form
\beq
\eta -\eta'\simeq -\fr{\lam_{\tau}^2R_{\tau }}{16\pi^2}\fr{b\al_2}{|b\al_2+e^{i\phi}|^2}\sin \phi ~.
\la{eta-eta1}
\eeq
Note once again that the phase $\eta -\eta'$, determining the lepton asymmetry, is proportional to $\sin \phi$, which itself is related to the
phase $\de $ of the lepton mixing matrix. The model gives the relation between them by Eq. (\ref{rel2-A}). Also, it is rather impressive that other
parameters, $b$ and $\al_2$, appearing in (\ref{eta-eta1}) can be calculated by the lepton mixing matrix elements through the relations
(\ref{rel1-A}), (\ref{defA1234}).
\begin{figure}
\begin{center}
\leavevmode
\leavevmode
\vspace{2.5cm}
\includegraphics{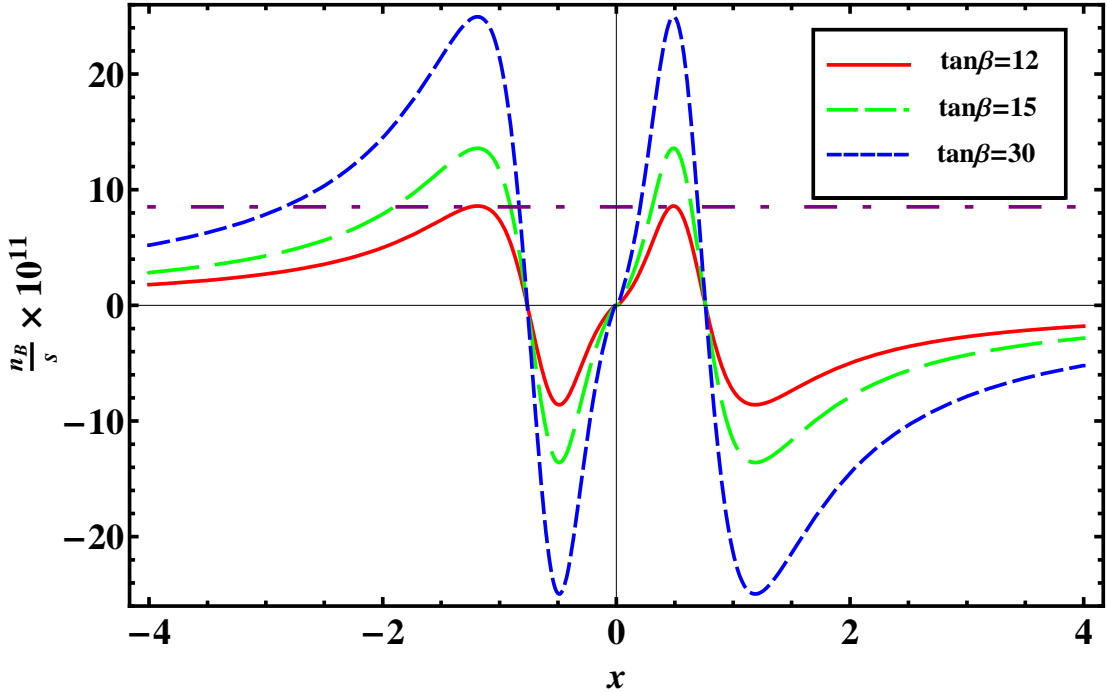}
\end{center}
\vs{5cm}
\caption{Baryon asymmetry for normal hierarchical case (texture A), for different values of $\tan \bt $ and $M=10^7$~GeV, $\de =1.3$.
}
\label{f:NH-tanbeta}
\end{figure}

The masses of two right handed neutrinos are
\beq
M_1=|M|(1-\ka )~,~~~~M_2=|M|(1+\ka )~,~~~~~{\rm with}~~~~\ka\simeq \left | x(b\al_2+e^{i\phi})\right |\fr{\bar{\bt}^2}{4\pi^2}\ln \fr{M_G}{M}~.
\la{N-masses}
\eeq
Here, the unknown parameter $x$ appears which is free and can be varied upon numerical calculations. Finally, we give expressions build from the elements
of the matrix $(\hat{Y}_{\nu}^{\dag}\hat{Y}_{\nu})$ appearing in the expressions of the CP asymmetries of Eq. (\ref{res-lept-asym}). These are:
$$
(\hat{Y}_{\nu}^{\dag}\hat{Y}_{\nu})_{11}\simeq \fr{\bar{\bt}^2}{2}\l x^2(1+\al_1^2+\al_2^2)+1+b^2+2x|b\al_2+e^{i\phi}|\r ~,
$$
$$
(\hat{Y}_{\nu}^{\dag}\hat{Y}_{\nu})_{22}\simeq \fr{\bar{\bt}^2}{2}\l x^2(1+\al_1^2+\al_2^2)+1+b^2-2x|b\al_2+e^{i\phi}|\r ~,
$$
\beq
{\rm Im}(\hat{Y}_{\nu}^{\dag}\hat{Y}_{\nu})_{21}^2\simeq \fr{\lam_{\tau}^2R_{\tau }}{16\pi^2}
\bar{\bt}^4xb\al_2\l x^2(1+\al_1^2+\al_2^2)-1-b^2\r \fr{\sin \phi }{|b\al_2+e^{i\phi}|} ~.
\la{hatY2-elements}
\eeq
In order to compute generated baryon asymmetry of the Universe, recall that the  lepton asymmetry is converted to the baryon asymmetry via sphaleron processes
\cite{Kuzmin:1985mm}  and is given by $\frac{n_B}{s}\simeq -1.48\times10^{-3}({\kappa_f}^{(1)}\epsilon_1+{\kappa_f}^{(2)}\epsilon_2) $,
where the efficiency factors ${\kappa_f}^{(1,2)}$  are given by the extrapolating formula \cite{Giudice:2003jh}:
\begin{equation}
{\kappa_f}^{(1,2)}=\left(\frac{3.3\times 10^{-3}{\rm eV}}{{\tilde{m}}_{1,\,2}}+\left(\frac{{\tilde{m}}_{1,\,2}}
{0.55\times 10^{-3}{\rm eV}}\right)^{1.16}\right)^{-1},\nonumber\\
\end{equation}
\begin{equation}
{\rm with}\,\,\, {\tilde{m}}_1=\frac{(v\sin \bt)^2}{M_1}(\hat{Y}_{\nu}^{\dag}{\hat{Y_{\nu}}})_{11}\,\,, \,\,\,\,
\,\,{\tilde{m}}_2=\frac{(v\sin \bt)^2}{M_2}(\hat{Y}_{\nu}^{\dag}{\hat{Y_{\nu}}})_{22}~.
\end{equation}
Collecting all this, we can now calculate $\fr{n_B}{s}$. One can try the different values of $M$ in a mass range which would not
cause the gravitino problem. We can also try different values of the phase $\de $, relevant also for the CP violation into the
neutrino oscillations. As we have already mentioned, there is one more free parameter $x$, which we will vary. It is quite interesting that this
system, by requiring to have  baryon asymmetry in the range of the observed amount $\l \fr{n_B}{s}\r_{\rm exp}=(8.75\pm 0.23)\cdot 10^{-11}$,
dictates the preferred range for the MSSM parameter $\tan \bt $.
The reason for this is simple. The strength of the Yukawa coupling $\lam_{\tau }$, determining the amount of the CP violation [see Eq. (\ref{eta-eta1})],
depends on the value of $\tan \bt $: $\lam_{\tau}=\fr{m_{\tau}}{v}\sq{1+\tan^2\bt}$. By simple but quite complete numerical simulation we
obtain, in this model, the low bound on the $\tan \bt$. Upon calculations we take into account the renormalization effects. Namely,
the running of $\lam_{\tau }$.
Obtained low bound for $\tan \bt$ is: $\tan \bt\stackrel{>}{_\sim }12$ (corresponds to $|\de |\simeq 1.3$ and $M=10^7$~GeV, $R_{\tau }=0.617$).
Smaller values of $\tan \bt $ do not give sufficient baryon asymmetry.
This also indicates that the non SUSY version
(i.e. SM augmented by two RHNs) of this scenario will fail to generate baryon asymmetry through the leptogenesis.
The presented scenario also allows to derive the low bound for the absolute value of the phase $\de $. This comes out from the maximal allowed value
of $\tan \bt \stackrel{<}{_\sim }58$ (from the requirement that $\lam_{b,\tau }\stackrel{<}{_\sim }1$ all the way up to the GUT scale).
 With $\tan \bt =58$, $M=10^7$~GeV
($R_{\tau }=2.17$) in order to have needed baryon asymmetry we should have $|\de |\stackrel{>}{_\sim }0.012$.
It is interesting to note that for $\tan \bt \stackrel{<}{_\sim }35$, for generating the baryon asymmetry we need
$|\de |\stackrel{>}{_\sim }0.1$. This limit for the CP violating phase is within the reach of future experiments.
In Figs. \ref{f:NH-tanbeta} and \ref{f:NH-M}  we plot $\fr{n_B}{s}$ for different choices of the model parameters.

\begin{figure}
\begin{center}
\leavevmode
\leavevmode
\vspace{2.5cm}
\includegraphics{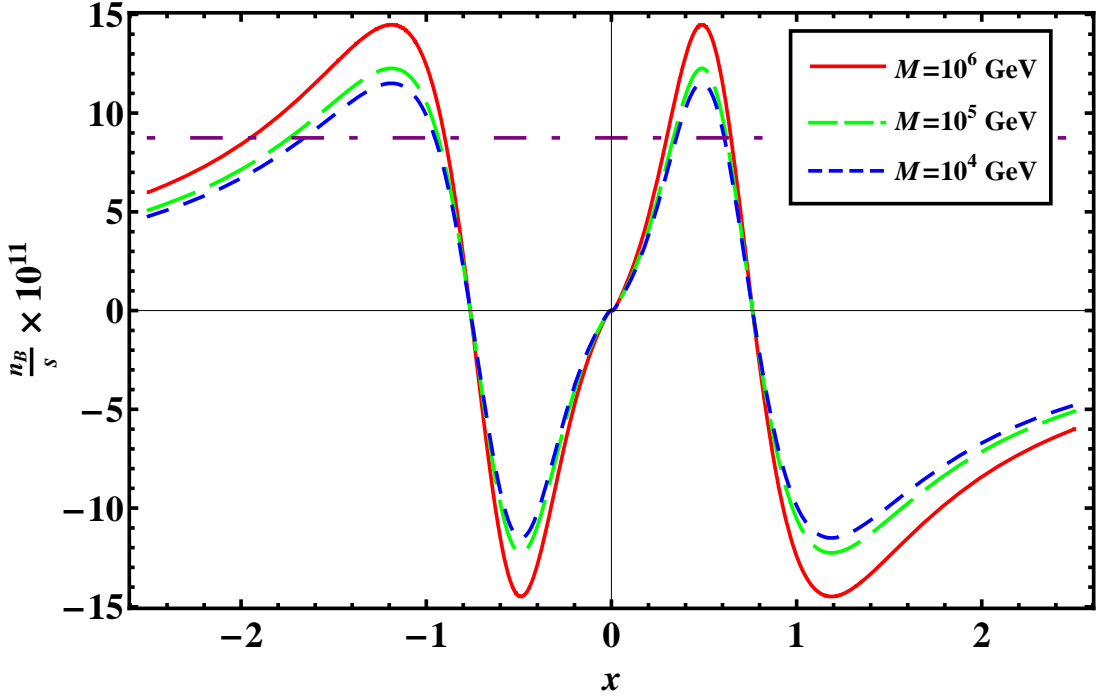}
\end{center}
\vs{5cm}
\caption{Baryon asymmetry for normal hierarchical case (texture A), for different values of $M$ and $\tan \bt =15$, $\de=1.3$.
}
\label{f:NH-M}
\end{figure}

\begin{center}
{\bf Leptogenesis for Inverted Hierarchical Case}
\end{center}

Now we study the leptogenesis for the inverted hierarchical case. We note right away that the scenario with texture $B_2$ of (\ref{textureB12})
does not work for the leptogenesis. The reason is following. Due to the zero in $(3,2)$ entry of this texture, it is easy to see
from Eq. (\ref{sol-2loop}) that the $\lam_{\tau }$ coupling do not contribute to the CP asymmetry induced at 2-loop level.
The couplings $\lam_e$ and $\lam_{\mu }$ do contribute, however they are small and can not induce needed asymmetry.

Thus, we focus here only on case with texture $B_1$. For this case, it is convenient to write $Y_{\nu }$ with the parameterization
\beq
{\rm Texture~ B_1:}~~~~Y_{\nu}=\l \!\begin{array}{cc}
x\al_1&b\\
x\al_2&0\\
 xe^{i\phi}&1
  \end{array}\!\r \!\cdot \!\bar{\bt} ~,
\la{1textureB12}
\eeq
where, as in case of texture A, by suitable phase redefinition of $l_{1,2,3}, N_{1,2}$ superfields we left with  only single phase $\phi $.
 Remaining parameters are real.
First we will express the model parameters $\al_{1,2}, b, \bar{\bt }$ in terms of matrix elements of $U$, neutrino mass $m_2$, the $M$, and $x$.
By solving the equations derived from the relation (\ref{relM-Md})
we obtain
$$
{\rm For~ Texture~B_1}~:~~~~\al_1=\left | \fr{{\cal B}_2{\cal B}_4}{{\cal B}_1{\cal B}_3}\right |~,~~~\al_2=2\left | \fr{{\cal B}_2}{{\cal B}_1}\right |~,~~
b=\left | \fr{{\cal B}_3}{{\cal B}_2}\right |~,~~~
\bar{\bt }=\fr{1}{v\sin \bt}\l \fr{m_2}{2}\left | \fr{{\cal B}_1M}{x}\right |\r^{1/2}~,
$$
\beq
\phi ={\rm Arg}\l \fr{{\cal B}_2^2{\cal B}_4}{{\cal B}_1{\cal B}_3^2}\r ~,
\la{rels-B2}
\eeq
where
\beq
{\cal B}_1=U_{\tau 2}^2-U_{\tau 1}^2\fr{U_{\mu 2}^2}{U_{\mu 1}^2}~,~~~
{\cal B}_2=U_{\mu 2}U_{\tau 2}-U_{\tau 1}\fr{U_{\mu 2}^2}{U_{\mu 1}}~,~~~
{\cal B}_3=U_{e2}U_{\mu 2}-U_{e1}\fr{U_{\mu 2}^2}{U_{\mu 1}}~,~~~
{\cal B}_4=U_{e2}^2-U_{e1}^2\fr{U_{\mu 2}^2}{U_{\mu 1}^2}~.
\la{defB1234}
\eeq
As we see, also in this case the phase $\phi $ is related to the $\de $-phase of the leptonic mixing matrix $U$ (see Eq. (\ref{Ustan})).
In particular using the relation (\ref{predB1}) in (\ref{rels-B2}) for $\phi $ and performing simple algebra we derive
\beq
\phi \simeq {\rm Arg}\l \fr{\De m_{\rm sol}^2}{|\De m_{\rm atm}^2|}\fr{e^{i\de }}{4s_{13}\tan \te_{23}}-1 \r ~~~
\Longrightarrow ~~~~~\phi \simeq \pi -\fr{\De m_{\rm sol}^2}{|\De m_{\rm atm}^2|}
\fr{\cot \te_{23}}{4\sin \te_{13}}\sin \de ~.
\la{phi-delta-ratioB1}
\eeq
Since the phase $\phi $ will appear in the leptonic CP asymmetry, with relation (\ref{phi-delta-ratioB1}) we will be able
to make calculations in terms of measured neutrino oscillation  parameters and the CP phase $\de $.
In Fig. \ref{f:phi-del} the correlation between $\phi $ and $\de $ is shown.

\begin{figure}
\begin{center}
\leavevmode
\leavevmode
\vspace{2.5cm}
\includegraphics{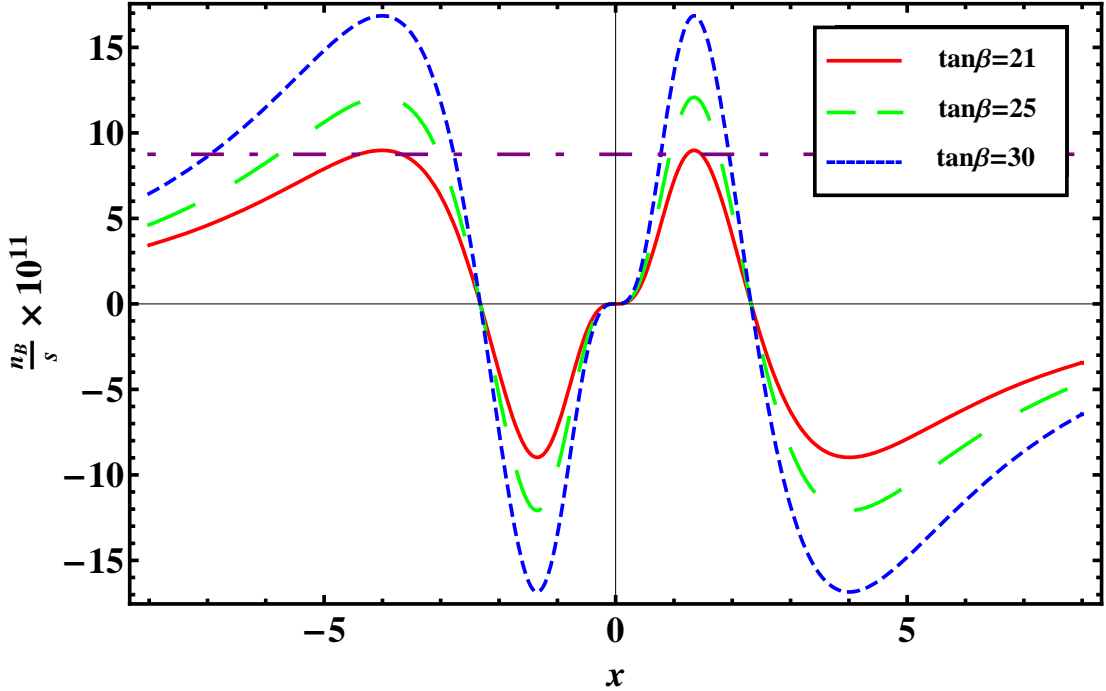}
\end{center}
\vs{5cm}
\caption{Baryon asymmetry for inverted hierarchical case (texture $B_1$), for different values of $\tan \bt $ and $M=10^4$~GeV, $\de=0.96$.
}
\label{f:IH-tanbeta}
\end{figure}

Now we are ready to investigate the leptogenesis for the inverted hierarchical scenario ($B_1$).
The way of calculation is same as was presented in the previous subsection, so we will
keep discussion short and give only several expressions and final results. Using the expressions of Eqs. (\ref{sol-1loop})-(\ref{renormMN})
and the form of the texture $B_1$ in (\ref{1textureB12}), for the phase mismatch (in analogy of Eq. (\ref{eta-eta1}) we obtain
\beq
\l \eta -\eta'\r^{(B_1)}\simeq -\fr{\lam_{\tau}^2R_{\tau }}{16\pi^2}\fr{b\al_1}{|b\al_1+e^{i\phi}|^2}\sin \phi ~,
\la{eta-eta1B}
\eeq
where here and below we will use superscript `$(B_1)$' in order to distinguish expressions corresponding to the scenario $B_1$
from those of the texture $A$.
Moreover, for the splitting parameter (in analog to Eq. (\ref{N-masses})) we have
\beq
\ka^{(B_1)}\simeq \left | x(b\al_1+e^{i\phi})\right |\fr{\bar{\bt}^2}{4\pi^2}\ln \fr{M_G}{M}~.
\la{kap-splB}
\eeq
We will also give the expression for ${\rm Im}(\hat{Y}_{\nu}^{\dag}\hat{Y}_{\nu})_{21}^2$ which will help to understand
some physics. We have
\beq
\l {\rm Im}(\hat{Y}_{\nu}^{\dag}\hat{Y}_{\nu})_{21}^2\r^{(B_1)}\simeq \fr{\lam_{\tau}^2R_{\tau }}
{16\pi^2}\bar{\bt}^4xb\al_1\l x^2(1+\al_1^2+\al_2^2)-1-b^2\r \fr{\sin \phi}{|b\al_1+e^{i\phi}|} ~.
\la{hatY2-elementsB1}
\eeq
Note, that according to (\ref{phi-delta-ratioB1}) the phase $\phi $ is close to $\pi $ and one may suspect that also final result
for the CP violation should be suppressed by the factor$\sim \De m_{\rm sol}^2/(|\De m_{\rm atm}^2|4\sin \te_{13})\approx 1/20$.
However, such suppression do not takes place because the combination  $|b\al_1+e^{i\phi}|$, appearing in the denominator
of the last multiplier of (\ref{hatY2-elementsB1}), is suppressed by precisely same factor! Indeed, using
the relations of Eqs. (\ref{predB1}), (\ref{rels-B2})-(\ref{phi-delta-ratioB1}) we derive
\beq
\left |b\al_1+e^{i\phi}\right |\simeq \fr{\De m_{\rm sol}^2}{|\De m_{\rm atm}^2|}
\fr{\cot \te_{23}}{4\sin \te_{13}}~.
\la{absAl1bpl1}
\eeq
With these for the combination appearing in (\ref{hatY2-elementsB1}) we get
\beq
\fr{\sin \phi}{|b\al_1+e^{i\phi}|}\simeq \sin \de ,
\la{enhanceCP}
\eeq
showing that suppression factors mentioned above drop out and it is maximized with $|\de |\simeq 1.115$ (maximal allowed
value which is acceptable for viable neutrino sector).
Moreover, because of the suppression of the combination  $\left |b\al_1+e^{i\phi}\right |$, also the RHN mass splitting
parameter in (\ref{kap-splB}) gets additional suppression, which makes two RHNs more degenerate. This also gives some enhancement
of the (resonant) CP asymmetry factors $\ep_{1,2}$.

\begin{figure}
\begin{center}
\leavevmode
\leavevmode
\vspace{2.5cm}
\includegraphics{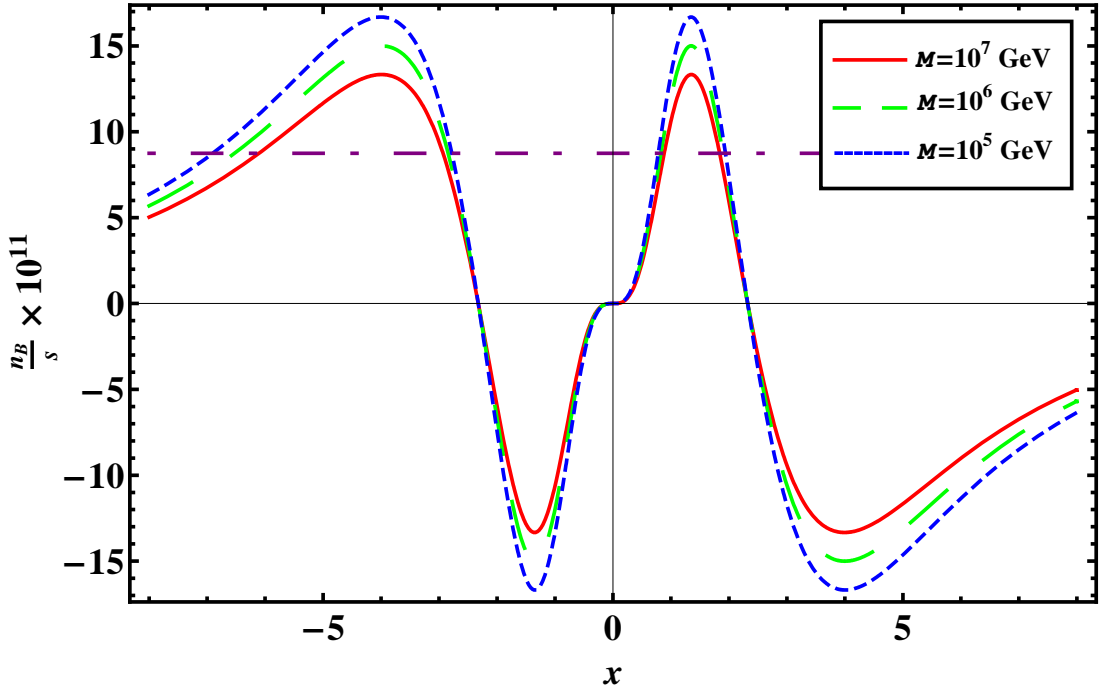}
\end{center}
\vs{5cm}
\caption{Baryon asymmetry for inverted hierarchical case (texture $B_1$), for different values of $M$ and $\tan \bt =30$, $\de=0.96$.
}
\label{f:IH-M}
\end{figure}

Without bothering to give other expressions, we will move to the presentation of the main results. In this scenario, from the requirement
of needed baryon asymmetry, the $\tan \bt $ is bounded from below.
Interesting thing is that the leptogenesis dictates $\tan \bt \stackrel{>}{_\sim } 21$ (lower values do not give sufficient baryon asymmetry)
For obtaining this low bound we have taken $M=10^4$~GeV ($R_{\tau }\simeq 0.71$) and  maximal allowed value for the $\de \simeq 0.96$.
Note that within this scenario low values of $M$ give larger lepton asymmetries.
It is also possible to derive low bound for $|\de |$. This is obtained by largest allowed (from requirement $\lam_{b\tau }\stackrel{<}{_\sim }1$
up to the GUT scale) value of $\tan \bt $. Namely, with $\tan \bt \simeq 58$, $M=10^4$~GeV (we have for these choices $R_{\tau }\simeq 1.95$), needed baryon asymmetry
can be generated with $|\de |\stackrel{>}{_\sim }0.021$ (note that for $|\de |=0.021$, for acceptable solar mixing angle we should choose
$\sin^2\te_{23}\simeq 0.6$ and $\sin^2\te_{13}\simeq 0.04$).
Worthwhile for noting that for $\tan \bt \stackrel{<}{_\sim }45$ for generating sufficient baryon asymmetry we need $|\de |\stackrel{>}{_\sim }0.1$.
The latter value is within the reach of planned experiments.
We have performed numerical calculations without approximations and made sure that our analytical expressions, presented above,
are good approximations.
In Figs. \ref{f:IH-tanbeta} and  \ref{f:IH-M} we show baryon asymmetries for several different choices of the model parameters.

%
%

%
%
%
%
%

\section{Summary}

In this paper we have considered an extension of MSSM with two quasi-degenerate right--handed neutrinos. Our motivation
was to realize resonant leptogenesis which avoids the gravitino problem generic for low scale SUSY scenarios. With this
setup we have classified all viable texture zeros of the neutrino Dirac Yukawa matrices which lead to
consitent predictions. We find three
predictive scenarios, each with one texture zero. One model has normal hierarchical neutrino mass spectrum, while the remaining two have inverted hierarchical mass pattern.
The predictive power of these models show up also in the resonant leptogenesis. The model with the normal mass hierarchy (texture $A$)
and one of the inverted hierarchical scenario (with texture $B_1$) lead to the successful leptogenesis.
In Appendix we have discussed the impact of the soft SUSY breaking terms on the CP asymmetry generated by RH sneutrino decays and concluded that with natural choice of the soft SUSY breaking terms, scalar RH neutrinos do not contribute sizably to the total baryon asymmetry.
Thus, the baryon asymmetry is due to fermionic RHN decays and the leptonic CP
phase is directly related to to CP violation in neutrino oscillation.
Putting together the predictions from the neutrino sector and the results from leptogenesis calculations, we have obtained the following predictions:

\vs{0.15cm}

\hs{-0.4cm}For normal hierarchical case (texture $A$)
$$
\sin^2\te_{13}\stackrel{>}{_\sim } 0.05~,~~~~~|\de |\stackrel{>}{_\sim } 0.012~,~~~~~
m_{\bt \bt }=0~,~~~~~\tan \bt \stackrel{>}{_\sim }12~;
$$
$$
{\rm with}~~~~\tan \bt \stackrel{<}{_\sim }35~,~~~~|\de |\stackrel{>}{_\sim } 0.1~.
$$

\vs{0.15cm}

\hs{-0.4cm}For the inverted hierarchical case corresponding to texture $B_1$:
$$
\te_{13}\stackrel{>}{_\sim}0.12 ~,~~~0.021\stackrel{<}{_\sim } |\de |\stackrel{<}{_\sim }0.96~,~~~
 0.013~{\rm eV} \stackrel{<}{_\sim }m_{\bt \bt} \stackrel{<}{_\sim }0.023~{\rm eV}~,~~~\tan \bt \stackrel{>}{_\sim }21~;
$$
$$
{\rm with}~~~~\tan \bt \stackrel{<}{_\sim }45~,~~~~|\de |\stackrel{>}{_\sim } 0.1~.
$$

\vs{0.15cm}

\hs{-0.4cm}The texture $B_2$ do not generate the baryon asymmetry within this scenario and other mechanism need to be invoked \cite{inprep}.
However, from the viewpoint of the neutrino sector the texture $B_2$ is viable and gives:
$$
\te_{13}\stackrel{>}{_\sim}0.129 ~,~~~~~~ |\pi -\de |\stackrel{<}{_\sim }0.91~,~~~
 0.013~{\rm eV} \stackrel{<}{_\sim }m_{\bt \bt} \stackrel{<}{_\sim }0.023~{\rm eV}~.
$$

Future experiments will examine the viability of these scenarios.

\section*{\small Acknowledgement}
\vs{-0.2cm}  The work is supported in part by DOE grant
DE-FG02-04ER41306 and DE-FG02-ER46140. Z.T. is also partially
supported by GNSF grant 07\_462\_4-270.

\section*{Appendix: Asymmetry Via $\tl{N}$ Decays }

In this appendix we will discuss the contribution to the net baryon asymmetry from the out of equilibrium resonant decays of the right
handed sneutrinos (RHS). With inclusion of the soft SUSY breaking terms, the RHS mass spectrum and couplings will be altered and
 one should expect result different from that corresponding to the fermionic RHN decays. Besides soft SUSY breaking couplings, there are
other particularities, highlighted below, which distinguish cases of RHN and RHS decays.
We are considering the system with two RHN superfields $N_{1,2}$ which have two complex scalar components $\tl{N}_{1,2}$. With SUSY breaking term, the masses of RHS's will differ from their fermionic partners' masses. Thus we will have four real mass-eigenstate RHS's $\tl{n}_{i=1,2,3,4}$
with masses $\tl{M}_{i=1,2,3,4}$ respectively. Assuming that the SUSY scale is smaller (at least by factor of 10) than the scale $M$
(the overall tree level mass  for the RHN superfields) we expect that the states $\tl{n}_i$ remain quasi-degenerate. To study
the resonant $\tl{n}$-decays we will apply ressumed effective amplitude technic \cite{Pilaftsis:1997jf}. An effective amplitudes
for the real $\tl{n}_i$ decay, say into the lepton $l_{\al }$ ($\al=1,2,3$ is a generation index) and antilepton $\ov{l}_{\al }$ respectively are given by \cite{Pilaftsis:1997jf}
\beq
\hat{S}_{\al i}=S_{\al i}-\sum_jS_{\al j}\fr{\Pi_{ji}(\tl{M}_i)(1-\de_{ij})}{\tl{M}_i^2-\tl{M}_j^2+\Pi_{jj}(\tl{M}_i)}~,~~
\hat{\ov{S}}_{\al i}=S_{\al i}^*-\sum_jS_{\al j}^*\fr{\Pi_{ji}(\tl{M}_i)(1-\de_{ij})}{\tl{M}_i^2-\tl{M}_j^2+\Pi_{jj}(\tl{M}_i)}~,
\la{eff-amps}
\eeq
where $S_{\al i}$ is a tree level amplitude and $\Pi_{ij}$ is a two point Green function's (polarization operator of $\tl{n}_i-\tl{n}_j$)
absorptive part. The CP asymmetry is then given by
\beq
\ep_i^{sc}=\fr{\sum_{\al }\l |\hat{S}_{\al i}|^2-|\hat{\ov{S}}_{\al i}|^2\r}{\sum_{\al }\l |\hat{S}_{\al i}|^2+|\hat{\ov{S}}_{\al i}|^2\r }~.
\la{asym-byS}
\eeq
We will apply (\ref{eff-amps}) and (\ref{asym-byS}) for our scenario, however, also derive general expressions applicable for different models.

Toegether with superpotential couplings (\ref{Wlept}) we include the following soft SUSY breaking terms
\beq
V_{\rm SB}^{\nu }=\tl{l}A_{\nu }\tl{N}h_u-\fr{1}{2}\tl{N}^TB_N\tl{N}+{\rm h.c.}
\la{soft-br}
\eeq
We do not display here soft mass$^2$ terms, such as $\tl{m}_{1,2}^2|\tl{N}_{1,2}|^2$, because $B_N$ plays much more significant
role in the splitting of RHS masses. For simplicity we will assume at GUT scale ($M_G$)  the `proportionality'  $A_{\nu }\propto Y_{\nu }$
and degeneracy in $B_N\propto M_N$. Thus,
\beq
{\rm at}~~\mu =M_{G}:~~~~~A_{\nu }=m_AY_{\nu }~,~~~~~~B_N=m_BM_N~.
\la{prop-GUT}
\eeq
Similarly, for the charged lepton sector we can assume $A_{e}=m_EY_{e}$. Performing RG studies, similar way as we have done in section
\ref{s:res}, we will have
\beq
{\rm at}~~\mu =M:~~~~~B_N \simeq m_BM \l \begin{array}{cc}- (1+2\fr{m_A}{m_B}) \delta_{N}&1\\1&-(1+2\fr{m_A}{m_B}) \delta^*_N\end{array}\r~.
\la{effectiveBN}
\eeq
Note that with $A_{\nu }=m_AY_{\nu }$ and $A_{e}=m_EY_{e}$ at high scale, the $A_{\nu }$ will remain well aligned with $Y_{\nu }$
also at low scales. With diagonalization of total mass matrix of the RHS's, for  mass-eigenstate ($\tl{n}_i$) masses we get
$$
\tl{M}_1^2=|M|^2(1-|\de_N|)^2-|M||m_B-(m_B+2m_A)|\de_N|| ~,
$$
$$
\tl{M}_2^2=|M|^2(1-|\de_N|)^2+|M| | m_B-(m_B+2m_A)|\de_N|| ~,
$$
$$
\tl{M}_3^2=|M|^2(1+|\de_N|)^2-|M| | m_B+(m_B+2m_A)|\de_N|| ~,
$$
\beq
\tl{M}_4^2=|M|^2(1+|\de_N|)^2+|M| | m_B+(m_B+2m_A)|\de_N||~.
\la{tl-n-masses}
\eeq
Interaction of $\tl{n}$ states with leptons and sleptons has the form
\beq
\tl{h}_ulY_F\tl{n}+h_u\tl{l}Y_B\tl{n}+{\rm h.c.}
\la{}
\eeq
where
$$
Y_F=Y_{\nu }\tl{V}~,~~~Y_B=Y_{\nu }M_N^*\tl{V}^*+A_{\nu }\tl{V}~,{\rm where}~~~\tl{V}=U_N\l \rho_u e^{i\tl{\te }},~\rho_d \r ~,
$$
\beq
{\rm with}~~~
\rho_u=\fr{1}{\sq{2}}\l  \begin{array}{cc} 1&i\\0&0 \end{array}\r ~,~~~
\rho_d=\fr{1}{\sq{2}}\l  \begin{array}{cc} 0&0\\1&i \end{array}\r ~,~~~
\tl{\te }\simeq 2{\rm Im}\l \fr{m_A}{m_B}\r |\de_N|~.
\la{couplings}
\eeq
With these we can calculate the absorptive part of the polarization diagram with external legs $\tl{n}_i$ and $\tl{n}_j$. At 1-loop
level it is given by
\beq
\Pi_{ij}(p)=\fr{i}{8\pi}\l p^2Y_F^\dag Y_F+p^2Y_F^T Y_F^*+Y_B^\dag Y_B+Y_B^TY_B^*\r_{ij} ~,
\la{abs-part}
\eeq
where $p$ denotes external momentum  in the diagram.

Now we are ready to calculate the lepton asymmetry. Note that in unbroken SUSY limit, neglecting finite temperature effects ($T\to 0$),
the $\tl{N}$ decay does not produce lepton asymmetry. The reason for this is following. The decay of $\tl{N}$ in two fermion is
 $\tl{N}\to l\tl{h}_u$, while in two scalars is $\tl{N}\to \tl{l}^*h_u^*$. Since the rates of these processes are same due to SUSY  (at $T=0$),
 the lepton asymmetries created from these decays cancel each other.
 However, with  $T\neq 0$ the cancelation is partial and one has
\beq
\tl{\ep }_i=\ep_i(\tl{n}_i\to l\tl{h}_u)\De_{BF}~,
\la{ep-non-zero}
\eeq
with temperature dependent factor $\De_{BF}$ given in \cite{D'Ambrosio:2003wy}.
We note that Eq. (\ref{ep-non-zero}) is valid when we have the alignment $A_{\nu }=m_AY_{\nu }$. Without this alignment other terms
in r.h.s of  (\ref{ep-non-zero}) proportional to $m_A/M$ will appear. Since we are assuming the alignment and $m_A/M\stackrel{<}{_\sim }0.1$,
the SUSY breaking effects would not affect decay amplitudes significantly and we can apply (\ref{ep-non-zero}) for our study.
Thus, we just need to compute $\ep_i(\tl{n}_i\to l\tl{h}_u)$ - the asymmetry created by $\tl{n}_i$ decays in two fermions.
Using in (\ref{eff-amps}) $S_{\al i}=(Y_F)_{\al i}$, with (\ref{asym-byS}) after straightforward
calculation we obtain
$$
\ep_i(\tl{n}_i\to l\tl{h}_u) \simeq \fr{1}{(Y_F^\dag Y_F)_{ii}}
\left \{  2\sum_j\fr{(\tl{M}_i^2-\tl{M}_j^2){\rm Im}(\Pi_{ji})}{(\tl{M}_i^2-\tl{M}_j^2)^2+|\Pi_{jj}|^2}{\rm Im}(Y_F^\dag Y_F)_{ji}\right. +
$$
\beq
\left. \sum_{j,\hs{0.5mm}k}{\rm Im}(Y_F^\dag Y_F)_{kj}
\fr{(\tl{M}_j^2-\tl{M}_i^2){\rm Im}(\Pi_{kk})-(\tl{M}_k^2-\tl{M}_i^2){\rm Im}(\Pi_{jj})}
{\l (\tl{M}_i^2-\tl{M}_j^2)^2+|\Pi_{jj}|^2\r \l (\tl{M}_i^2-\tl{M}_k^2)^2+|\Pi_{kk}|^2\r }\Pi_{ji}\Pi_{ki}\right \}~.
\la{tiln-eps}
\eeq
In (\ref{tiln-eps}) for the absorptive part $\Pi $ we should use (\ref{abs-part}) with $p=\tl{M}_i$.
Now, the baryon asymmetry created from the lepton asymmetry due to $\tl{n}$ decays is:
\beq
\fr{\tl{n}_B}{s}\simeq -8.46\cdot 10^{-4}\sum_{i=1}^4\fr{\tl{\ep }_i}{\De_{BF}}\eta_i=
-8.46\cdot 10^{-4}\sum_{i=1}^4\ep_i(\tl{n}_i\to l\tl{h}_u)\eta_i~,
\la{sc-asym}
\eeq
where we have taken into account that an effective number of degrees of freedom, including two RHN superfields, is
$g_*=228.75$. $\eta_i$ are an efficiency factors which depend on $\tl{m}_i\simeq \fr{(v\sin \bt )^2}{M}2(Y_F^\dag Y_F)_{ii}$, and take into account temperature effects by integrating the Boltzmann equations \cite{D'Ambrosio:2003wy}. Before discussing this in more details, it is more instructive
to see what are the effects of the soft SUSY breaking terms in the CP asymmetry given by Eq. (\ref{tiln-eps}).
The parameter $\ep_i$ is controlled by the imaginary parts of the elements of the matrix $Y_F^\dag Y_F$. First note that the phase $\tl{\te }$
appearing in this matrix (see Eq. (\ref{couplings})) for $M\stackrel{<}{_\sim }10^7$~GeV is $\tl{\te }\stackrel{<}{_\sim }10^{-10}$ and can be safely
ignored. With this, the matrix $Y_F^\dag Y_F$ has the form
\beq
Y_F^\dag Y_F=\l  \begin{array}{cc} \hat{\si }(\hat{Y}_{\nu }^\dag \hat{Y}_{\nu })_{11}&\hat{\si }(\hat{Y}_{\nu }^\dag \hat{Y}_{\nu })_{12}
\\\hat{\si }(\hat{Y}_{\nu }^\dag \hat{Y}_{\nu })_{21}&\hat{\si }(\hat{Y}_{\nu }^\dag \hat{Y}_{\nu })_{22} \end{array}\r ~,~~~
{\rm with}~~~\hat{\si }=\fr{1}{2}\l  \begin{array}{cc} 1&i\\-i&1 \end{array}\r ~,
\la{YFYF-form}
\eeq
where $\hat{Y}_{\nu }=Y_{\nu }U_N$ is the same matrix appearing in the CP asymmetries (\ref{res-lept-asym}) induced by fermionic RHN decays.
Note that the matrix $\hat{\si }$ has purely imaginary entries and they can be new sources for the CP violation. For instance, the element
$(Y_F^\dag Y_F)_{12}$ has the large phase. This means that there happens the `conversion' between $\tl{n}_1$ and $\tl{n}_2$ states. On the other hand,
from Eq. (\ref{tl-n-masses}) one can see that the degeneracy of $\tl{M}_1^2$ and $\tl{M}_2^2$ is split by the $B$-term and unless
$m_B\stackrel{<}{_\sim}10$~MeV the resonant enhancement does not happen (similar to the case of soft leptogenesis \cite{D'Ambrosio:2003wy}). Since the natural
value of $m_B$ is from few$\tm 100$~GeV to few TeV, we conclude that this channel does not give important contribution to the CP asymmetry. For those states amongst which degeneracy is not ruined (the `pairs' $\tl{n}_1-\tl{n}_3$ and $\tl{n}_2-\tl{n}_4$) by the $B$-terms, the CP asymmetry
is controlled not by imaginary components of $\hat{\si }$ but by ${\rm Im}(\hat{Y}_{\nu }^\dag \hat{Y}_{\nu })_{12}$ (like to those corresponding
to the fermionic RHN decays, Eq. (\ref{res-lept-asym})). Thus, the CP asymmetry via $\tl{n}_i$ decays would not be larger than asymmetry generated due
to their fermionic partners. Moreover, due to the efficiency factors $\eta_i$, the $\tl{n}_B/s$ turns out to get additional suppression in comparison
to the $n_B/s$ (the total baryon asymmetry due fermionic RHNs). We have checked this on two examples corresponding to the textures of
$A$ and $B_1$.
Namely, we have performed calculations for $(m_A,~m_B)=(10^3i,~10^3)$~GeV and for several choice of model
parameters $(\tan \bt , M, \de )$. For a given set of these parameters, for  fixed $x$ we can calculate
the values of the masses $\tl{m}_i=\fr{(w\sin \bt )^2}{M}2(Y_F^\dag Y_F)_{ii}$. With given values
of $\tl{m}_i$, according to Ref. \cite{D'Ambrosio:2003wy} we picked up the corresponding values of $\eta_i$ and with help of
Eqs. (\ref{tiln-eps}), (\ref{sc-asym}) calculated $\tl{n}_B/s$. For the texture $A$ we obtained $\fr{\tl{n}_B}{n_B}<4\cdot 10^{-3}$,
while for the texture $B_1$:  $\fr{\tl{n}_B}{n_B}<10^{-2}$.
These confirm that the baryon asymmetry via $\tl{n}$ decays is a negligible effect.
For completeness we also examined the case corresponding to texture $B_2$. The latter does not give relevant asymmetry also through $\tl{n}$ decays.

\end{document}